\documentclass[12pt]{elsart}

\usepackage{amssymb}
\usepackage[dvips]{graphicx}
\usepackage{latexsym}
\usepackage{amssymb}
\usepackage{cases}
\usepackage[compress]{cite}
\usepackage{xcolor}
\usepackage{enumerate}

\newtheorem{Theorem}{Theorem}
\newtheorem{defi}{Definition}

\journal{Physica A}

\begin{document}

\begin{frontmatter}

\title{Tutte polynomial of pseudofractal scale-free web}
\author[lable1,label2,label3]{Junhao Peng}
\ead{pengjh@gzhu.edu.cn}
\author[label2]{Guoai Xu}
\address[lable1]{School of Mathematics and Information Science, Guangzhou University , Guangzhou 510006 ,  China.}
\address[label2]{State Key Laboratory of Networking and Switching Technology,
Beijing University of Posts and Telecommunications , Beijing 100876 ,China.}
\address[label3]{Key Laboratory of Mathematics and Interdisciplinary Sciences of Guangdong
Higher Education Institutes, Guangzhou University,Guangzhou 510006 ,China.}
\begin{abstract}
The Tutte polynomial of a graph is a 2-variable polynomial which is quite important in both combinatorics and statistical physics. It contains various numerical  invariants and polynomial invariants  ,such as the number of  spanning trees,the number of spanning forests , the number of acyclic orientations , the reliability polynomial,chromatic polynomial and flow polynomial . In this paper,we study and gain recursive formulas for the Tutte polynomial of pseudofractal scale-free web(PSW) which implies logarithmic complexity algorithm is obtained to calculate the  Tutte polynomial of PSW  although it is NP-hard for general graph.We also obtain the rigorous solution  for the the number of spanning trees of PSW by solving the recurrence relations derived from Tutte polynomial ,which  give an alternative approach for explicitly determining the number of  spanning trees of PSW.Further more,we  analysis the all-terminal reliability of PSW and compare the results with that of  Sierpinski gasket which has the same number of nodes and edges with PSW. In contrast with the well-known conclusion that  scale-free networks are more robust against removal of nodes than homogeneous networks (e.g., exponential networks and regular networks).Our results show that  Sierpinski gasket (which is a regular network) are  more robust against random edge failures than  PSW (which is a scale-free network) .Whether it is true for any  regular networks and scale-free networks ,is still a unresolved problem.
\end{abstract}
\end{frontmatter}
\section{Introduction}
The Tutte polynomial is a two-variable polynomial which can be associated with a graph, a matrix, or, more generally, with a matroid . This polynomial was introduced by W.T. Tutte \cite{tutte47, tutte54, tutte67} and has many interesting applications in several areas of sciences such as  combinatorics, probability, statistical mechanics and computer science. It is quite interesting since several combinatorial, enumerative and algebraic properties of the graph can be investigated by considering special evaluations of it\cite{Monaghan08}. For instance, one gets information about the number of spanning trees\cite{chang07,zhang12}, spanning connected subgraphs\cite{chang09}, spanning forests\cite{chang08} and acyclic orientations\cite{chang10} of the graph by evaluating tutte polynomial  at particular points $(x, y)$  . Moreover, the Tutte polynomial contains several other polynomial invariants, such as flow polynomial\cite{chang03a} , reliability polynomial\cite{chang03b} and  chromatic polynomial\cite{tutte54,Rocek97,Shrock12}. It has also many interesting connections with statistical mechanical models such as the Potts model \cite{welsh00,Sokal05,Sokal11} and the percolation \cite{Oxley79}.
 Despite its ubiquity,there are no widely-available effective computational tools  to compute the Tutte polynomials of a general graph of reasonable size.It is shown that, many of the relevant coefficients do not even have good randomised approximation schemes and various decision problems based on the coefficients are NP-hard\cite{JVW90,Oxley02}.Although it is hard to to compute the Tutte polynomials,a lot of efforts have been devoted to the study the Tutte polynomials of different graph such as polygon chain graphs \cite{Sokal11}, Sierpinski gaskets\cite{Donno10}  and strips of lattices \cite{Chang08b,Chang04,Salas02,Shrock00}.However, polygon chain ,lattices and Sierpinski gaskets cannot well mimic the real-life networks, which have been recently found to synchronously exhibit two striking properties: scale-free behavior\cite{Barabasi99} and small-world effects \cite{Watts98}.

 Pseudofractal scale-free web(PSW) we studied is  a deterministically growing network introduced by S.N. Dorogovtsev\cite{Dorogovtsev02} which is used to model scale-free network with small-world effect.Lots of job was devoted to study its properties ,such as degree distribution ,degree correlation , clustering coefficient\cite{Dorogovtsev02,zhang07b} ,diameter\cite{zhang07b},average path length\cite{zhang07}, the number of spanning trees\cite{zhang10} and mean first-passage time for random walk \cite{zhang09}.As for its tutte polynomial and reliability polynomial ,to the best of our knowledge, related research was rarely reported .

  In this paper,we study and gain recursive formulas for the Tutte polynomial of PSW .The analytic method is based on the its recursive construction and self-similar structure .Recursive formulas for various invariants of Tutte polynomial can also obtained based on their connections with the Tutte polynomial .
  We also obtain the rigorous solution  for the the number of spanning trees by solving the recurrence relations derived from  the Tutte polynomial,which  coincides with the result obtained in \cite{zhang10} .Thus we give an alternative approach for explicitly determining the number of  spanning trees of PSW .Further more,we  analysis the all-terminal reliability of PSW and compare the results with that of  Sierpinski gasket which has the same number of nodes and edges with PSW. In contrast with the well-known conclusion that  scale-free networks are more robust against random node failures than homogeneous networks (e.g., exponential networks and regular networks) .Our results show  that  Sierpinski gasket (which is a regular network) are  more robust against random edge failures than  PSW (which is a scale-free network) .
\section{Preliminaries}\label{Section preliminare}

\subsection{The Tutte polynomial}\label{sec2.1}
Let $G=(V(G),E(G))$ denotes a graph with vertex set $V(G)$ and edge set $E(G)$; we will often write $V$ and $E$, when
there is no risk of confusion, and so $G=(V,E)$.  A subgraph $H=(V(H),E(H))$ of a graph $G=(V(G),E(G))$ is said \emph{spanning}
if the condition $V(H)=V(G)$ is satisfied. In particular, a \emph{spanning tree} of $G$ is a spanning subgraph of $G$ which
is a tree.Let $H$ be a spanning subgraph of $G$ and $k(H)$ be the number of connected components of $H$.then the rank $r(H)$ and the nullity $n(H)$ of $H$ are defined as
$$ r(H)=|V(H)| -k(H)= |V(H)|-k(H)$$
$$ n(H)=|E(H)| - r(H)=|E(H)|-|V(H)| + k(H)$$
\begin{defi}\label{tuttpoly}
Let $G=(V,E)$ be a graph. The Tutte polynomial $T(G;x,y)$ of $G$
is defined as\cite{Monaghan08}
\begin{equation}\label{DefTuPoly}
T(G;x,y)= \sum_{H\subseteq G}(x-1)^{r(G)-r(H)}(y-1)^{n(H)}
\end{equation}
where the sum runs over all the spanning subgraphs $H$ of $G$ .
\end{defi}\par
The Tutte polynomial can be evaluated at particular points $(x, y)$ to give numerical invariants,such as the number of spanning trees,
the number of forests and the number of connected spanning subgraphs.The following theorem\cite{Monaghan08} depicts the relations between the Tutte polynomial and its numerical invariants.
\begin{Theorem}\label{theorem1}
Let $G=(V,E)$ be a connected graph ,Then:
\begin{enumerate}
\item $T(G; 1,1)$ is the number of spanning trees;
\item $T(G; 1,2)$ is the number of spanning connected subgraphs of $G$;
\item $T(G; 2,1)$ is the number of spanning forests  of $G$;
\end{enumerate}
\end{Theorem} \par
The Tutte polynomial also has a variety of single-variable polynomial invariants associated with the graph,such as  the all-terminal reliability polynomial,flow polynomial and the chromatic polynomial.In this paper we only study the all-terminal reliability polynomial. Let $G=(V,E)$ be a graph,each edge of G has a known probability $p$ of being operational; otherwise it is failed. Operations of different edges are statistically independent,while the nodes of $G$ never fail.
The all-terminal reliability polynomial $R(G, p)$ of $G$ is defined as is the probability that there is a path of operational edges between any pair of vertices of $G$.The connection between the Tutte polynomial and all-terminal reliability polynomial is given by the following  theorem\cite{Monaghan08}.
\begin{Theorem} \label{theorem2}
 Let $G=(V,E)$ be a graph, Then
 \begin{equation}
R(G,p)=p^{|V(G)|-1}(1-p)^{|E(G)|-|V(G)|+1}T\left(G;1,\frac{1}{1-p}\right)
\label{PolyRel}
\end{equation}
\end{Theorem}
\subsection{Structure of PSW}
\label{sec2.2}
\begin{figure}
\begin{center}
\includegraphics[scale=0.5]{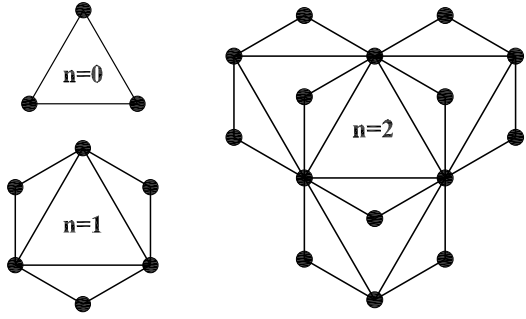}
\caption{Growth process for PSW from n=0 to n=2 }
\label{fig:1}       
\end{center}
\end{figure}

\begin{figure}
\begin{center}
\includegraphics[scale=0.6]{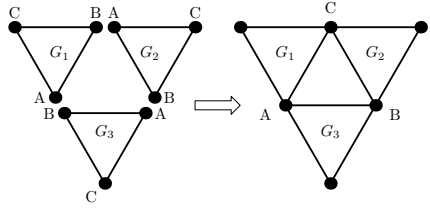}
\caption{Second construction method of PSW: $G(n+1)$ is composed of three copies of $G(n)$ denoted as  $G_1$,$G_2$,$G_3$ ,the three hub nodes of which  are represented by A,B,C in the corresponding triangle(The left side of the figure ).In the merging process,hub node A of $G_1$ and hub node B of $G_3$ ,hub node A of $G_3$ and hub node B of $G_2$ , hub node A of $G_2$ and hub node B of $G_1$ are identified as a hub node A,B,C of $G(n+1)$ respectively(The right side of the figure ).}
\label{fig:2}       
\end{center}
\end{figure}

The scale-free network  we studied is  a deterministically growing network which can be constructed iteratively \cite{Dorogovtsev02} .
We denote the pseudofractal scale-free web(PSW) after n iterations by $G(n)$ with $n\geq 0 $.
Then it is constructed as follows: For $n=0$, $G_0$ is a triangle. For $n\geq 1$, $G_{n}$ is
obtained from $G_{n-1}$: every existing edge in $G_{n-1}$ introduces a new node connected to both ends of the edge. The construction process of the first three generation is shown in Fig. \ref{fig:1}.
The PSW exhibits some typical properties of real networks. Its degree distribution $P(k)$ obeys a power law $P(k)\sim
k^{1+\ln 3/\ln 2}$\cite{Dorogovtsev02}, the average path length scales logarithmically with network order \cite{zhang07}. The network also has an equivalent  construction method\cite{zhang10,Bobe05},as can be seen in Fig. \ref{fig:2} : to obtain $G(n+1)$, one can make three copies of $G(n)$ and join them  at the three most connected nodes denoted by A,B,C, which are called the hub nodes of $G(n+1)$ in this paper.
  According to the second construction algorithm, one can see that at each step n, the total number of edges in the network increases by a factor of $3$. Thus,  the total number of edges for $G(n)$ is $E_n=3^{n+1}$.We can also find that the total number of nodes for $G(n)$ is $V_n=\frac{3^{n+1}+3}{2}$.
\section{Tutte polynomial of PSW }
\label{sec:2}
 Let us simply denote by $T_n(x,y)$ the Tutte polynomial of PSW  $G(n)$  .In this section we study and gain recursive formulas for  $T_n(x,y)$.The analytic method is based on the  the relation between spanning subgraphs of $G(n+1)$ and spanning subgraphs of $G(n)$.According to the second construction algorithm,we find that there exists a bijection between spanning subgraphs of $G(n+1)$ and spanning subgraphs of  $G_1, G_2,G_3$  inside $G(n+1)$,while the subgraphs $G_1, G_2, G_3$  are isomorphic to $G(n)$.Indeed,if H is a spanning subgraph of $G(n+1)$ ,we can
uniquely determines three spanning subgraphs $H_1$, $H_2$ and $H_3$  of $G_1, G_2$ and $G_3$ ; viceversa, given three spanning subgraphs $H_1$, $H_2$ and $H_3$ of $G_1, G_2$ and $G_3$, respectively, then their union provides a spanning subgraph $H$ of the whole $G(n+1)$. Therefore, according to the Definition \ref{tuttpoly},the Tutte polynomial  of PSW $G(n+1)$ can be rewritten as
 \begin{equation}
T_{n+1}(x,y) = \sum_{H_i\subseteq G_i, i=1,2,3
}(x-1)^{r(G(n+1))-r(H)}(y-1)^{n(H)},
\label{TuPoly2}
\end{equation}
where $H_i$ is the spanning subgraph of $G_i$ ,and H is the union of $H_1$, $H_2$ and $H_3$.In order to obtain the recursive formulas for  $T_n(x,y)$, we want to know the relations between $r(G(n+1))$ and $r(G(n))$ ,$r(H)$ and $r(H_i)$, $n(H)$ and $n(H_i)$, for $i=1,2,3$.
It is easy to know that $r(G(n+1))=3r(G(n))-1$ and $|V(H)|=|V(H_1)|+|V(H_2)|+|V(H_3)|-3$,
$|E(H)|=|E(H_1)|+|E(H_2)|+|E(H_3)|$ ,for every spanning subgraph $H$ of  $G(n+1)$.
Furthermore, two possibilities can occur.\par
 1)In the  spanning subgraph $H_i$ of $G_i$,  the  hub nodes of $G({n+1})$ belong to the same connected component,for any $i,1\leq i\leq 3$, then
$$
k(H) = k(H_1) +k(H_2) + k(H_3) -2 \qquad \mbox{and} \qquad r(A) =
r(H_1) + r(H_2) + r(H_3) -1
$$
Thus
$$n(H)=|E(H)|-r(H)=n(H_1) + n(H_2) + n(H_3) + 1$$
$$
r(G(n+1))-r(H) = \sum_{i=1}^3 (r(G(n))-r(H_i))
$$
Hence,  one gets:
\begin{eqnarray}
&&(x-1)^{r(G(n+1))-r(H)}(y-1)^{n(H)} \nonumber \\
&=&(y-1) \prod_{i=1}^3(x-1)^{r(G(n))-r(H_i)}(y-1)^{n(H_i)}
\label{Tucas1}
\end{eqnarray}

2)For certain $i,1\leq i\leq 3$ , in the  spanning subgraph $H_i$,  the hub nodes of $G({n+1})$  do not belong to
the same connected component,we have
$$
k(H) = k(H_1) +k(H_2) + k(H_3) -3 \qquad \mbox{and} \qquad r(A) =
r(H_1) + r(H_2) + r(H_3)
$$
Moreover
$$
n(H)= n(H_1) + n(H_2) + n(H_3)
$$
Hence
$$
r(G(n+1))-r(H) = \sum_{i=1}^3 (r(G(n))-r(H_i))-1$$

Thus
\begin{eqnarray}
&&(x-1)^{r(G(n+1))-r(H)}(y-1)^{n(H)}\nonumber \\
&=& \frac{1}{(x-1)} \prod_{i=1}^3   (x-1)^{r(G(n))-r(H_i)}(y-1)^{n(H_i)}
\label{Tucas2}
\end{eqnarray}
Let $D_n$ denotes the set of spanning subgraphs of $G(n)$,we define the
following partition on $D_n$ :
\begin{itemize}
\item $D_{1,n}$ denotes the set of spanning subgraphs of $G(n)$,
where the three hub nodes belong to the same connected
component;
\item $D_{2,n}^C$ denotes the set of spanning subgraphs of $G(n)$, where the hub nodes A and B
 belong to the same connected component, and C belongs to a different component. Similarly, $D_{2,n}^A$ (or  $D_{2,n}^B$ )denotes the set
of spanning subgraphs of $G(n)$, where A(or B )does not belong to the  connected
component containing the other two hub nodes;
\item $D_{3,n}$ denotes the set of spanning subgraphs of $G(n)$, where the three hub nodes belong to three different connected components.
\end{itemize}

Thus, for any $n\geq 0$, we have
\begin{equation}\label{partition}
D_n=D_{1,n} \cup  D_{2,n}^A \cup  D_{2,n}^B \cup  D_{2,n}^C \cup D_{3,n}
\end{equation}
For  $n\geq 0$,let us define  polynomial of $x,y$
\begin{eqnarray}  \label{t1n}
 T_{1,n}(x,y)= \sum_{H\in D_{1,n}}(x-1)^{r(\Gamma_n)-r(H)}(y-1)^{n(H)}\nonumber
\end{eqnarray}
Similarly,we can also define  polynomials :$T_{2,n}^A(x,y),T_{2,n}^B(x,y),T_{2,n}^C(x,y),T_{3,n}(x,y)$,
while the sum is conducted on $D_{2,n}^A, D_{2,n}^B, D_{2,n}^C,D_{3,n}$ respectively.For symmetry,we have
$$
T_{2,n}^A(x,y) =  T_{2,n}^B(x,y) = T_{2,n}^C(x,y),
$$
and  we can simply use 
$T_{2,n}(x,y)$ to denote one of the three polynomials. According to Definition
\ref{tuttpoly} , we have:
\begin{equation}\label{tn1}
T_n(x,y) = T_{1,n}(x,y) + 3T_{2,n}(x,y) + T_{3,n}(x,y)
\end{equation}
Furthermore,we obtain the following theorem based on the  relation between spanning subgraphs of $G(n+1)$ and spanning subgraphs of $G(n)$.
\begin{Theorem}\label{TuPloyPSW}
For  $n\geq 0$, the Tutte polynomial $T_n(x,y)$ of $G(n)$ is given by
\begin{eqnarray}\label{tn2}
T_n(x,y)=T_{1,n}(x,y)+3(x-1)P_{n}(x,y)+(x-1)^2Q_{n}(x,y)
\end{eqnarray}
where  $T_{1,n}(x,y)$, $P_{n}(x,y)$,$Q_{n}(x,y) $ satisfy the following recurrence relation:
\begin{eqnarray}\label{t1ns}
T_{1,n+1}(x,y) &=& (y-1)T_{1,n}^3 +3(y-1)(x-1)T_{1,n}^2P_{n} \nonumber \\
              &&+3(x-1)^2(y-1)T_{1,n}N_{n}^2+(x-1)^3(y-1)P_{n}^3    \nonumber \\
              &&+3(x-1)T_{1,n}^2Q_{n}+6T_{1,n}^2P_{n}+12(x-1)T_{1,n}P_{n}^2 \nonumber \\
              &&+6(x-1)^2P_{n}^3 +3(x-1)^3P_{n}^2Q_{n}+6(x-1)^2T_{1,n}P_{n}Q_{n}
\end{eqnarray}
\begin{eqnarray}\label{pn}
P_{n+1}(x,y)&=&4T_{1,n}P_{n}^2+4(x-1)T_{1,n}P_{n}Q_{n}+4(x-1)P_{n}^3 \nonumber \\
& &+4(x-1)^2P_{n}^2Q_{n}+(x-1)^2T_{1,n}Q_{n}^2+(x-1)^3P_{n}Q_{n}^2
\end{eqnarray}
\begin{eqnarray}\label{qn}
Q_{n+1}(x,y)&=&8P_{n}^3+ 12(x-1)P_{n}^2Q_{n}+6(x-1)^2P_{n}Q_{n}^2+(x-1)^3Q_{n}^3
\end{eqnarray}
with initial conditions
$$
T_{1,0}(x,y)=y+2 \qquad P_0(x,y)=Q_0(x,y)=1.
$$
and $T_{1,n}$,$P_{n}$,$Q_{n} $ is shorthand of $T_{1,n}(x,y)$, $P_{n}(x,y)$,$Q_{n}(x,y) $ respectively.
\end{Theorem}
{\bf Proof}:
We find that  $x-1$ divides $T_{2,n}(x,y)$ and $(x-1)^2$
divides $T_{3,n}(x,y)$  for any $n\geq 0$ which will be proved later.As a consequence, we can write
\begin{eqnarray}\label{semplify}
T_{2,n}(x,y) = (x-1)P_n(x,y) \nonumber \\
T_{3,n}(x,y) = (x-1)^2Q_n(x,y)
\end{eqnarray}
where $P_n(x,y)$ and $ Q_n(x,y)$ are polynomials of $x,y$.
Thus we obtain Eq.(\ref{tn2}) from Eq.(\ref{tn1}).\par
Now we will proof $T_{1,n}(x,y)$, $T_{2,n}(x,y)$,$T_{2,n}(x,y) $ satisfy the following recurrence relations which lead to the results of the theorem.
\begin{eqnarray}\label{t1nr}
T_{1,n+1}(x,y) &=& (y-1)T_{1,n}^3 +3(y-1)T_{1,n}^2T_{2,n}+3(y-1)T_{1,n}T_{2,n}^2 \nonumber \\
              & &+(y-1)T_{2,n}^3 +\frac{1}{x-1} ( 3T_{1,n}^2T_{3,n}+6T_{1,n}^2T_{2,n}   \nonumber \\
              & &+12T_{1,n}T_{2,n}^2+6T_{2,n}^3 +3T_{3,n}T_{2,n}^2+6T_{1,n}T_{2,n}T_{3,n} )
\end{eqnarray}
\begin{eqnarray}\label{t2nr}
T_{2,n+1}(x,y)&=&\frac{1}{x-1}(4T_{1,n}T_{2,n}^2+4T_{1,n}T_{2,n}T_{3,n}+4T_{2,n}^3 \nonumber\\
& &+4T_{2,n}^2T_{3,n}+T_{1,n}T_{3,n}^2+T_{2,n}T_{3,n}^2)
\end{eqnarray}
\begin{eqnarray}\label{t3nr}
T_{3,n+1}(x,y)&=&\frac{1}{x-1}\left(8T_{2,n}^3+ 12T_{2,n}^2T_{3,n}+6T_{2,n}T_{3,n}^2+T_{3,n}^3\right)
\end{eqnarray}
with initial conditions
$$
T_{1,0}(x,y)=y+2 \qquad T_{2,0}(x,y)=x-1 \qquad T_{3,0}(x,y)=(x-1)^2.
$$
and $T_{1,n}$,$T_{2,n}$,$T_{3,n} $ is shorthand of $T_{1,n}(x,y)$, $T_{2,n}(x,y)$,$T_{3,n}(x,y) $ respectively.\par
 The strategy of the proof  is to study all the possible configurations of spanning subgraphs $H_i$ in the three copies $G_i$ of $G(n)$ inside $G(n+1)$, for $i=1,2,3$, and analyze which kind of contribution they give to $T_{1,n+1}(x,y)$, $T_{2,n+1}(x,y)$ and $T_{3,n+1}(x,y)$.\par
\begin{figure}
\begin{center}
\includegraphics[scale=0.5]{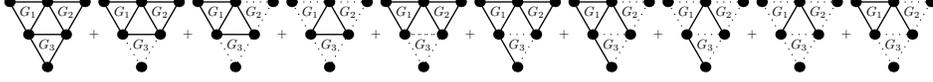}
\caption{The  possible configurations of spanning subgraphs $H_i(i=1,2,3)$ for $T_{1,n+1}(x,y)$.The two hub nodes of $G_i$ are connected by a solid line if they are in in the same connected component, and connected by a dotted line if they are not in the same connected component.}
\label{fig:3}
\end{center}
\end{figure}

For $T_{1,n+1}(x,y)$,we find it has $10$  possible configurations of spanning subgraphs $H_i$( $i=1,2,3$) which is shown in Fig. \ref{fig:3}.In the first configuration, $H_i\in D_{1,n}$ for any $i=1,2,3$. This contributes to $T_{1,n+1}(x,y)$ by a term $(y-1)T_{1,n}^3$ according to Eq.(\ref{Tucas1}), since  in the  spanning subgraph $H_i$ of $G_i$,  the  hub nodes of $G_{n+1}$ belong to the same connected component,for any $i,1\leq i\leq 3$.In the second configuration,$H_i\in D_{1,n}$ holds for two $i\in \{1,2,3\}$ and $H_i\in D_{2,n}^C$ holds for the last $i$(for example,$H_1\in D_{1,n}$,$H_2\in D_{1,n}$,$H_3\in D_{2,n}^C$).This contributes to $T_{1,n+1}(x,y)$ by a term $3(y-1)T_{1,n}^2T_{2,n}$ according to Eq.(\ref{Tucas1}). Computing  the contributions to $T_{1,n+1}(x,y)$ of the $10$  possible configurations and adding them together ,we obtain Eq.(\ref{t1nr}).\par
\begin{figure}
\begin{center}
\includegraphics[scale=0.5]{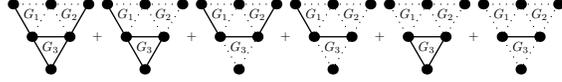}
\caption{The  possible configurations of spanning subgraphs $H_i(i=1,2,3)$ for $T_{2,n+1}^C(x,y)$.The two hub nodes of $G_i$ are connected by a solid line if they are in in the same connected component, and connected by a dotted line if they are not in the same connected component.}
\label{fig:4}
\end{center}
\end{figure}

 For $T_{2,n+1}(x,y)$, we study $T_{2,n+1}^C(x,y)$ only by symmetry.We find it has $6$  possible configurations  which is shown in Fig. \ref{fig:4}.In the first configuration, $H_3\in D_{1,n}$ $H_1\in D_{2,n}^A\cup D_{2,n}^B$,$H_2\in D_{2,n}^A\cup D_{2,n}^B$. This contributes to $T_{2,n+1}^C(x,y)$ by a term $\frac{4}{x-1}T_{1,n}T_{2,n}^2$ according to Eq.(\ref{Tucas2}), since  in the  spanning subgraph $H_1$ or $H_2$ ,  the  hub nodes of $G_{n+1}$ do not belong to
the same connected component.In the second configuration,$H_3\in D_{1,n}$ $H_1\in D_{2,n}^A\cup D_{2,n}^B$,$H_2\in D_{3,n}$.This contributes to $T_{2,n+1}^C(x,y)$ by a term $\frac{4}{x-1}T_{1,n}T_{2,n}T_{3,n}$ according to Eq.(\ref{Tucas2}). Computing  the contributions to $T_{2,n+1}^C(x,y)$ of all the $6$  possible configurations and adding them together ,we obtain Eq.(\ref{t2nr}).\par
\begin{figure}
\begin{center}
\includegraphics[scale=0.5]{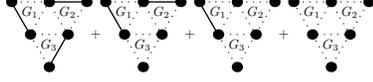}
\caption{The  possible configurations of spanning subgraphs $H_i(i=1,2,3)$ for $T_{3,n+1}(x,y)$.The two hub nodes of  $G_i$ are connected by a solid line if they are in in the same connected component, and connected by a dotted line if they are not in the same connected component.}
\label{fig:5}
\end{center}
\end{figure}

 For $T_{3,n+1}(x,y)$, we find it has $4$  possible configurations  which is shown in Fig. \ref{fig:5}.In the first configuration, $H_i\in D_{2,n}^A\cup D_{2,n}^B$,for any $i=1,2,3$.This contributes to $T_{3,n+1}(x,y)$ by a term $\frac{8}{x-1}T_{2,n}^3$ according to Eq.(\ref{Tucas2}), since  in the  spanning subgraph $H_1$  $H_2$ and $H_3$ ,  the  hub nodes of $G_{n+1}$ do not belong to the same connected component.In the second configuration,$H_i\in D_{2,n}^A\cup D_{2,n}^B$ holds for two $i\in \{1,2,3\}$ and $H_i\in D_{3,n}$ holds for the last $i$(for example,$H_1\in D_{2,n}^A\cup D_{2,n}^B$, $H_2\in D_{2,n}^A\cup D_{2,n}^B$,$H_3\in D_{3,n}$).This contributes to $T_{3,n+1}(x,y)$ by a term $\frac{12}{x-1}T_{2,n}^2T_{3,n}$. Computing  the contributions to $T_{3,n+1}(x,y)$ of all the $4$  possible configurations and adding them together ,we obtain Eq.(\ref{t3nr}).\par
For the  initial conditions,It is easy to verify  according to the definition.\par
Now,we come back to proof the recurrence relations that Eqs.(\ref{t1ns}),(\ref{pn}),(\ref{qn}) show.
First,we find $x-1$ divides $T_{2,n}(x,y)$ and $(x-1)^2$ divides $T_{3,n}(x,y)$,for  $n\geq 0$. The results can be proved  by mathematical induction based on the recurrence relations  that Eqs.(\ref{t2nr}),(\ref{t3nr}) show.\par
Inserting  Eq.(\ref{semplify}) into Eqs.(\ref{t1nr}),(\ref{t2nr}),(\ref{t3nr}) for $T_{2,n}(x,y)$ and $T_{3,n}(x,y)$,we obtain Eqs.(\ref{t1ns}),(\ref{pn}),(\ref{qn}).The initial conditions for $P_n(x,y)$ and $Q_n(x,y)$ is easy to verified according to the initial conditions for $T_{2,n}(x,y)$ and $T_{3,n}(x,y)$.\par
{\bf Remark:}Although it is NP-hard to calculate the Tutte polynomials for general graph,the recurrence relations we obtain shows that we can calculate the  Tutte polynomials for PSW with time complexity $O(n)=O(log(V_n))$.Thus ,we have obtain logarithmic complexity algorithm to calculate the  Tutte polynomial of PSW .\par
We can also obtain  the  recursive formulas for various invariants of Tutte polynomial based on their connections with the Tutte polynomial,such as the number of  spanning trees,the number of connected spanning subgraphs,the number of spanning forests,the number of acyclic orientations ,the reliability polynomial and the chromatic polynomial .In this paper ,we only study the number of  spanning trees and the reliability polynomial.
\section{Exact result for spanning trees of PSW}
\label{sec:3}
Let us denote by $N_{ST}(n)$ the number of  spanning trees of PSW $G(n)$ .According to Theorem \ref{theorem1} and Theorem \ref{TuPloyPSW},it is easy to know
   $$N_{ST}(n)=T_n(1,1)=T_{1,n}(1,1)$$
 Further more , for any $n\geq 0$,we have  the following recurrence relation
 $$N_{ST}(n+1)=6N_{ST}(n)^2P_n \qquad and \qquad P_{n+1}=4N_{ST}(n)P_n^2$$
 where $P_n$  is abbreviation of  $P_n(1,1) $  ,and the  initial conditions is
$$
N_{ST}(0)=T_{1,0}(1,1)=3 \qquad P_0=1.
$$
 Thus
  \begin{eqnarray}
  N_{ST}(n)&=&6N_{ST}(n-1)^2P_{n-1} \nonumber \\
         &=&6^{(1+2)}4^1[N_{ST}(n-2)]^{(2\times2+1)}P_{n-1}^{(2+2)} \nonumber \\
         &=&\ldots \nonumber \\
         &\triangleq&6^{a_k}4^{b_k}[N_{ST}(n-k)]^{c_k}P_{n-1}^{d_k}   \nonumber \\
         &=&\ldots \nonumber \\
         &\triangleq&6^{a_{n}}4^{b_{n}}[N_{ST}(0)]^{c_{n}}P_{0}^{d_{n}} \label{enst}
\end{eqnarray}
where $a_k,b_k,c_k,d_k, k>1$ satisfy the following recurrence relations .
 \begin{equation}
  a_{k}=a_{k-1}+c_{k-1}    \label{ak}
  \end{equation}
 \begin{equation}
  b_{k}=b_{k-1}+d_{k-1}   \label{bk}
  \end{equation}
 \begin{equation}
  c_{k}=2c_{k-1}+d_{k-1}  \label{ck}
  \end{equation}
 \begin{equation}
  d_{k}=c_{k-1}+2d_{k-1}  \label{dk}
\end{equation}
with initial conditions
$$
a_1 =1\qquad  b_1 =0 \qquad c_1=2 \qquad and \qquad d_1 =1
$$
Thus 
 \begin{equation}
  c_{k}+d_{k}=3(c_{k-1}+d_{k-1}) =3^{k-1}(c_1+d_1)=3^k \nonumber
   \end{equation}
 \begin{equation}
  c_{k}-d_{k}=c_{k-1}-d_{k-1} =c_1-d_1=1 \nonumber
\end{equation}
Then
\begin{equation}
  c_{k}=\frac{3^k+1}{2}   \label{eck}
\end{equation}
\begin{equation}
  d_{k}=\frac{3^k-1}{2}    \label{edk}
\end{equation}
Note
 $$ a_{k}-a_{k-1}=c_{k-1}    \qquad \mbox{and} \qquad
  b_{k}-b_{k-1}=d_{k-1} $$
 Thus
\begin{eqnarray}
  a_{k}-a_{1}=\sum_{i =1}^{k-1}(a_{i+1}-a_i)=\sum_{i =1}^{k-1}c_{i}=\frac{k-1}{2}+\frac{3^k-3}{4}    \nonumber\\
  b_{k}-b_{1}=\sum_{i =1}^{k-1}(b_{i+1}-b_i)=\sum_{i =1}^{k-1}d_{i}=-\frac{k-1}{2}+\frac{3^k-3}{4}    \nonumber
\end{eqnarray}
Hence
\begin{equation}
  a_{k}=\frac{k+1}{2}+\frac{3^k-3}{4}   \label{eak}
\end{equation}
\begin{equation}
  b_{k}=-\frac{k-1}{2}+\frac{3^k-3}{4}    \label{ebk}
\end{equation}
 Substituting Eqs.(\ref{eck}),(\ref{edk}), (\ref{eak}),(\ref{ebk})into Eq.(\ref{enst}),we  obtain the following result.
\begin{Theorem}\label{NumSPT}
For any $n\geq 0$,the number of  spanning trees of PSW $G(n)$ is given by
\begin{equation}\label{NST}
  N_{ST}(n)=2^{\frac{3^{n+1}-2n-3}{4}}3^{\frac{3^{n+1}+2n+1}{4}}
\end{equation}
\end{Theorem}
Note:The result  coincides with the result  obtained in \cite{zhang10},and we give an alternative approach for explicitly determining the number of  spanning trees for PSW.It is smaller than the the number of spanning trees for Sierpinski gasket\cite{ChangChenYang07}.
\section{Reliability analysis of PSW}
\label{sec:4}

\subsection{All-terminal reliability  of PSW}
\label{sec:4.1}
 In this section, we look upon PSW $G(n)$ as a probabilistic graph.Each edge of  $G(n)$ has a known probability $p$ of being operational; otherwise it is failed. Operations of different edges are statistically independent,while the nodes of $G(n)$ never fail.
The all-terminal reliability polynomial $R(G(n), p)$ of $G(n)$ is defined as is the probability that there is a path of operational edges between any pair of vertices of $G(n)$.
In general case ,the calculation of all-terminal reliability polynomial is NP-hard\cite{Ball80}. But for PSW,we obtain recursive formulas for all-terminal reliability polynomial which implies that logarithmic complexity algorithm is obtained.Further more,we get a approximate solution of $R(G(n), p)$ based on the recursive formulas it satisfy,which shows that all terminal reliability decreases approximately as a exponential function of network order.

let us simply denote by $T_{1,n}$,$P_{n}$,the expression $T_{1,n}(1,\frac{1}{1-p})$,$P_{n}\left(1,\frac{1}{1-p}\right)$ respectively for $n\geq 0$, we obtain the following recurrence relations  from Theorem \ref{TuPloyPSW}.
  \begin{equation}
 T_n\left(1,\frac{1}{1-p}\right)=T_{1,n}\left(1,\frac{1}{1-p}\right)
 \end{equation}
 \begin{equation}\label{ERT2}
T_{1,n+1}\left(1,\frac{1}{1-p}\right) =\frac{p}{1-p}T_{1,n}^3 +6T_{1,n}^2P_{n}
\end{equation}
\begin{equation}\label{ERN}
P_{n+1}\left(1,\frac{1}{1-p}\right)= 4T_{1,n}P_{n}^2
\end{equation}
with initial conditions$$
T_{1,0}\left(1,\frac{1}{1-p}\right) =\frac{3-2p}{1-p} \qquad P_0\left(1,\frac{1}{1-p}\right) = 1.
$$

For $n\geq 0$,let us simply denote by $R(n)$ the reliability polynomial $R(G(n),p)$ of PSW $G(n)$ ,and  define
 \begin{equation}
B(n)= p^{(V_{n}-2)}(1-p)^{(E_{n}-V_{n}+2)}P_{n}
\end{equation}
We  obtain the following recurrence relation from Theorem \ref{theorem2}
\begin{eqnarray}\label{RofPSW}
R(n+1) &=&p^{V_{n+1}-1}(1-p)^{E_{n+1}-V_{n+1}+1}T_{1,n+1}(1,\frac{1}{1-p}) \nonumber \\
    &=&p^{3V_{n}-4}(1-p)^{3E_{n}-3V_{n}+4}(\frac{p}{1-p}T_{1,n}^3+6T^2_{1,n}P_{n}) \nonumber \\
    &=&p^{3(V_{n}-1)}(1-p)^{3(E_{n}-V_{n}+1)}T_{1,n}^3 \nonumber \\
    & &+6p^{2(V_{n}-1)}(1-p)^{2(E_{n}-V_{n}+1)}T_{1,n}^2 \cdot p^{(V_{n}-2)}(1-p)^{(E_{n}-V_{n}+2)}P_{n} \nonumber \\
    &=&R(n)^3+6R(n)^2B(n)
\end{eqnarray}
and
\begin{eqnarray}
\label{BofPSW}
B(n+1)&=& p^{(V_{n+1}-2)}(1-p)^{(E_{n+1}-V_{n+1}+2)}P_{n+1} \nonumber \\
     &=&4 p^{(3V_{n}-5)}(1-p)^{(3E_{n}-3V_{n}+5)}T_{1,n}P_{n}^2 \nonumber \\
     &=&4R(n)B(n)^2
\end{eqnarray}
where $V_n$,$E_n$ denote the total number of nodes and edges of $G(n)$ respectively and the initial conditions is
$$R(0) =p^2(3-2p) \qquad     B(0) = p(1-p)^2 $$
Note that $R(n)\gg B(n)$ while $ n\rightarrow \infty$,
ones get
\begin{eqnarray}\label{PolyRR}
R(n+1)+2B(n+1)&=&R(n)^3+6R(n)^2B(n)+4R(n) B(n)^2 \nonumber \\
                &=&(R(n)+2B(n))^3-4(R(n)+2B(n))B(n)^2 \nonumber \\
                &\approx &(R(n)+2B(n))^3
\end{eqnarray}
 Thus 
\begin{eqnarray}\label{APoR}
R(n) & \approx & (R(n-1)+2B(n-1))^3  \nonumber \\
 & \approx &(R(1)+2B(1))^{3^{n-1}}=[p(2-p)]^{3^{n-1}} \approx [p(2-p)]^{\frac{2}{3}V_n}
\end{eqnarray}
 which shows that all terminal reliability decreases approximately as a exponential function of network order $V_n$.The reason  we don't use $R(n-1)^3$ as  a approximation  of $R(n)$ is that it has lager relative error than $(R(n-1)+2B(n-1))^3$.
 In fact, we  find that $B(n)$ is  the probability that $G(n)$ is split into two different connected components such that one of them contains the hub nodes A,B ,the other  one contains the hub node C  .Thus $R(n)+2B(n)<1$,for any $n\geq 0$.

\subsection{Comparison of all-terminal reliability between Sierpinski gasket and PSW}
The Sierpinski gasket is a fractal which can be constructed iteratively\cite{mandelbrot1982}. The starting point is a triangle. Divide its three sides in two segments of equal length. Connect the midpoints to get four inner triangles and paint the three external ones.Apply the same process to the inner triangles but the middle one.Sierpinski gasket is the limiting set for this construction.\par
\begin{figure}[htbp]
\begin{center}
\unitlength 0.9mm \hspace*{3mm}
\begin{picture}(60,20)
\put(0,0){\line(1,0){6}}
\put(0,0){\line(3,5){3}}
\put(6,0){\line(-3,5){3}}
\put(3,-4){\makebox(0,0){$SG(0)$}}
\put(12,0){\line(1,0){12}}
\put(12,0){\line(3,5){6}}
\put(24,0){\line(-3,5){6}}
\put(15,5){\line(1,0){6}}
\put(18,0){\line(3,5){3}}
\put(18,0){\line(-3,5){3}}
\put(18,-4){\makebox(0,0){$SG(1)$}}
\put(30,0){\line(1,0){24}}
\put(30,0){\line(3,5){12}}
\put(54,0){\line(-3,5){12}}
\put(36,10){\line(1,0){12}}
\put(42,0){\line(3,5){6}}
\put(42,0){\line(-3,5){6}}
\multiput(33,5)(12,0){2}{\line(1,0){6}}
\multiput(36,0)(12,0){2}{\line(3,5){3}}
\multiput(36,0)(12,0){2}{\line(-3,5){3}}
\put(39,15){\line(1,0){6}}
\put(42,10){\line(3,5){3}}
\put(42,10){\line(-3,5){3}}
\put(42,-4){\makebox(0,0){$SG(2)$}}
\end{picture}

\vspace*{5mm}
\caption{Growth process for  Sierpinski gasket $SG(n)$ from n=0 to n=2 .}
\label{sgfig}
\end{center}
\end{figure}
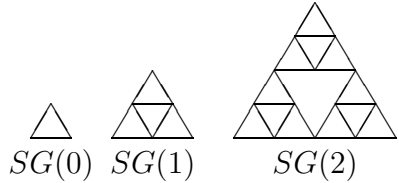

If we look upon the Sierpinski gasket as a network,it is a deterministically growing network which has the same starting point with PSW .But the method of iteration is different from PSW\cite{Hilfer1984}.
 The construction process of the first three generation is shown in Fig.\ref{sgfig}.We denote the Sierpinski gasket after n iterations by $SG(n)$ with $n\geq 0 $.It has the same number of nodes and edges with PSW for any $n\geq 0 $,but the structure is quite different.For Sierpinski gasket,except the $3$ outmost nodes which have degree $2$, all other vertices of $SG(n)$ have degree $4$. In the large $n$ limit, $SG(n)$ is $4$-regular.But PSW is a scale free netwokr whose  degree distribution obeys power law .Thus, Sierpinski gasket and PSW are typical examples of regular network and scale free network which have the same number of nodes and edges .\par
Alfredo Donno\cite{Donno10} found that for each $n\geq 0$, the all-terminal reliability polynomial of  Sierpinski gasket $SG(n)$ is
given by
$$
R(SG(n),p) =p^{V_{n}-1}(1-p)^{E_{n}-V_{n}+1}T_{1,n}(1,\frac{1}{1-p})
$$
with $T_n\left(1,\frac{1}{1-p}\right)=
T_{1,n}\left(1,\frac{1}{1-p}\right)$ and
\begin{equation}\label{olga2}
T_{1,n+1}\left(1,\frac{1}{1-p}\right) =\frac{p}{1-p}T_{1,n}^3
+6T_{1,n}^2N_{n}
\end{equation}
\begin{equation}\label{olga1}
N_{n+1}\left(1,\frac{1}{1-p}\right)=
\frac{p}{1-p}T_{1,n}^2N_{n}+T_{1,n}^2M_{n}+7T_{1,n}N_{n}^2
\end{equation}
\begin{equation}\label{olga0}
M_{n+1}\left(1,\frac{1}{1-p}\right)=\frac{3p}{1-p}T_{1,n}N_{n}^2+ 12T_{1,n}N_{n}M_{n}+14N_{n}^3
\end{equation}
with initial conditions
$$
T_{1,0}\left(1,\frac{1}{1-p}\right) =\frac{3-2p}{1-p} \qquad
N_0\left(1,\frac{1}{1-p}\right) = M_0\left(1,\frac{1}{1-p}\right)
=1.
$$
For $n\geq 0$,let us simply denote by $R_{s}(n)$ the reliability polynomial $R(SG(n),p)$  ,and  define
 \begin{equation}
B_s(n)= p^{(V_{n}-2)}(1-p)^{(E_{n}-V_{n}+2)}N_{n}
\end{equation}
 \begin{equation}
T_s(n)= p^{(V_{n}-3)}(1-p)^{(E_{n}-V_{n}+3)}M_{n}
\end{equation}
We obtain the following recurrence relation .
\begin{eqnarray} \label{RofSG}
R_{s}(n+1) &=&p^{V_{n+1}-1}(1-p)^{E_{n+1}-V_{n+1}+1}T_{1,n+1}(1,\frac{1}{1-p}) \nonumber \\
    &=&p^{3V_{n}-4}(1-p)^{3E_{n}-3V_{n}+4}(\frac{p}{1-p}T_{1,n}^3+6T^2_{1,n}N_{n}) \nonumber \\
    &=&R_{s}(n)^3+6R_{s}(n)^2B_s(n)
\end{eqnarray}
and
\begin{eqnarray}  \label{BofSG}
B_s(n+1)&=& p^{(V_{n+1}-2)}(1-p)^{(E_{n+1}-V_{n+1}+2)}N_{n+1} \nonumber \\
     &=& p^{(3V_{n}-5)}(1-p)^{(3E_{n}-3V_{n}+5)}(\frac{p}{1-p}T_{1,n}^2N_{n}+T_{1,n}^2M_{n}+7T_{1,n}N_{n}^2) \nonumber \\
     &=&R_{s}(n)^2B_s(n)+R_{s}(n)^2T_s(n)+7R_{s}(n)B_s(n)^2
\end{eqnarray}
\begin{eqnarray}
T_s(n+1)&=& p^{(V_{n+1}-3)}(1-p)^{(E_{n+1}-V_{n+1}+3)}M_{n+1} \nonumber \\
     &=& p^{(3V_{n}-6)}(1-p)^{(3E_{n}-3V_{n}+6)}(\frac{3p}{1-p}T_{1,n}N_{n}^2+ 12T_{1,n}N_{n}M_{n}+14N_{n}^3) \nonumber \\
     &=&3R_{s}(n)B_s(n)^2+12R_{s}(n)B_s(n)T_s(n)+14B_s(n)^3
\end{eqnarray}
with initial conditions
 $$R_{s}(0) =p^2(3-2p) \qquad    B_{s}(0) = p(1-p)^2  \qquad   T_{s}(0) = (1-p)^3$$ \par
\begin{figure}
\begin{center}
\includegraphics[scale=0.8]{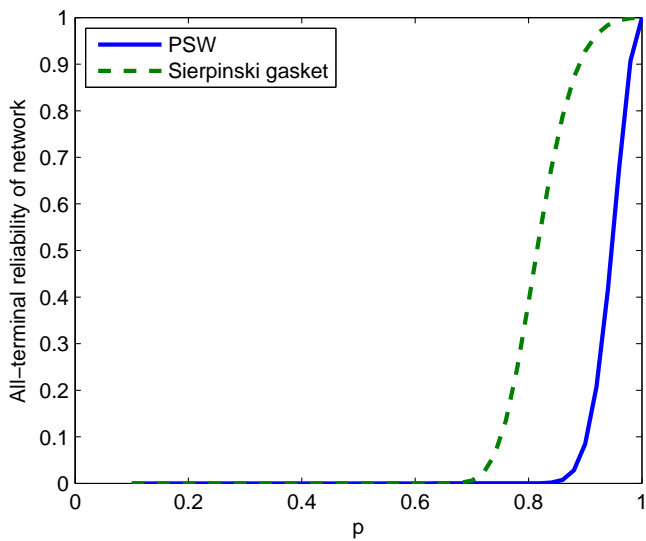}
\caption{The all-terminal reliability of Sierpinski gasket $SG(6)$ and PSW $G(6)$  obtained by direct calculation from Eqs.(\ref{RofPSW}) and (\ref{RofSG}) .}
\label{fig:7}
\end{center}
\end{figure}

Now,we compare all-terminal reliability between Sierpinski gasket and PSW while $p\in(0,1)$  .
For n=0,
$$R_{s}(0) =R(0) \qquad   B_{s}(0) =B(0)\qquad  $$
Thus $R_{s}(1) =R(1) $, But for  $n>0$,we can obtain from Eqs.(\ref{BofPSW}) and (\ref{BofSG}) that $B_{s}(n) >B(n)$. Thus ,for any $n>1$,we  find  from Eqs.(\ref{RofPSW} )and (\ref{RofSG})
\begin{equation}\label{ResultoCom}
R_{s}(n)>R(n)
\end{equation}
We  have calculated $R_{s}(n)$ and $R(n)$ for different $p\in(0,1)$ and $n$ by iteration.we find that $R_{s}(n)$ and $R(n)$ converge to 0 quickly with $n\rightarrow\infty$  and Eq.(\ref{ResultoCom}) holds for $n>1$.The results for $n=6$ are shown in Fig. \ref{fig:7}.Our results show that Sierpinski gasket is more robust than PSW against random edge failures.Thus,we obtain an example which shows that  regular networks(e.g.,Sierpinski gaskets) are  more robust than  scale-free networks(e.g. ,PSW)  against random edge failures.Whether it is true for any  regular networks and scale-free networks ,is still a unresolved problem. However, recent work~\cite{AlJeBa00,CaNeStWa00,CoErAvHa01} have shown that inhomogeneous networks, such as scale-free networks, are more robust than homogeneous networks (e.g., exponential networks and regular networks) with respect to random deletion of nodes. Thus, combining with our above result, we can reach the following  conclusion that networks (e.g.,scale-free networks) which  are more robust again random node failures  do not mean more robust again  random breakdown of edges than those (e.g., regular lattices) which are more vulnerable to random node failures .

\section{Conclusion}
\label{sec:5}
In this paper,we study and gain recursive formulas for the Tutte polynomial of PSW which implies that recursive formulas for various invariants of Tutte polynomial can also obtained based on their connections with the Tutte polynomial .
We also obtain the rigorous solution  for the the number of spanning trees of PSW by solving the recurrence relations derived from Tutte polynomial ,which  give an alternative approach for explicitly determining the number of  spanning trees of PSW.Further more,we  analysis the all-terminal reliability of PSW based on the the recurrence relations derived from Tutte polynomial and compare the result with that of  Sierpinski gasket. In contrast with the well-known conclusion that  scale-free networks are more robust than homogeneous networks (e.g., exponential networks and regular networks) with respect to random deletion of nodes.Our results show that there is an example  that regular networks(e.g.,Sierpinski gaskets) are  more robust than  scale-free networks(e.g.,PSW) against random edge failures.Whether it is true for any  regular networks and scale-free networks ,is still a unresolved problem.

\subsection*{Acknowledgment}

The authors are grateful to the anonymous referees for their valuable comments and suggestions. This work was supported  by the National High Technology Research and Development Program("863"Program) of China under Grant No. 2009AA01Z439.








\end{document}